\documentclass[preprint,english,aps,prl,superscriptaddress,showpacs]{revtex4}
\usepackage{graphicx}
\usepackage{babel}
\usepackage{amsmath}
\usepackage{amsfonts}
\usepackage{amssymb}

\begin{document}

\title{Hanbury-Brown Twiss correlations to probe the population statistics \\of GHz photons emitted by conductors} \author{J. Gabelli} \affiliation{Laboratoire Pierre Aigrain, D{\'e}partement de Physique de l'Ecole Normale Sup\'erieure, 24 rue Lhomond, 75231 Paris Cedex 05, France} \author{L.-H. Reydellet} \affiliation{Service de Physique de l'Etat Condens{\'e}, CEA Saclay, F-91191 Gif-sur-Yvette, France} \author{G. F{\`e}ve} \affiliation{Laboratoire Pierre Aigrain, D{\'e}partement de Physique de l'Ecole Normale Sup\'erieure, 24 rue Lhomond, 75231 Paris Cedex 05, France} \author{J.-M. Berroir} \affiliation{Laboratoire Pierre Aigrain, D{\'e}partement de Physique de l'Ecole Normale Sup\'erieure, 24 rue Lhomond, 75231 Paris Cedex 05, France} \author{B. Pla\c{c}ais} \affiliation{Laboratoire Pierre Aigrain, D{\'e}partement de Physique de l'Ecole Normale Sup\'erieure, 24 rue Lhomond, 75231 Paris Cedex 05, France} \author{P. Roche} \affiliation{Service de Physique de l'Etat Condens{\'e}, CEA Saclay, F-91191 Gif-sur-Yvette, France} \author{D.C. Glattli} \email{glattli@lpa.ens.fr} \affiliation{Laboratoire Pierre Aigrain, D{\'e}partement de Physique de l'Ecole Normale Sup\'erieure, 24 rue Lhomond, 75231 Paris Cedex 05, France} \affiliation{Service de Physique de l'Etat Condens{\'e}, CEA Saclay, F-91191 Gif-sur-Yvette, France}

\date{\today}

\begin{abstract}
We present the first study of the statistics of GHz photons in
quantum circuits, using Hanbury-Brown and Twiss correlations. The
superpoissonian and poissonian photon statistics of thermal and
coherent sources respectively made of a resistor and a
radiofrequency generator are measured down to the quantum regime
at milliKelvin temperatures. As photon correlations are linked to
the second and fourth moments of current fluctuations, this
experiment, which is based on current cryogenic electronics, may
become a standard for probing electron/photon statistics in
quantum conductors.

\end{abstract}
 \maketitle

\begin{figure}
\centerline{\includegraphics{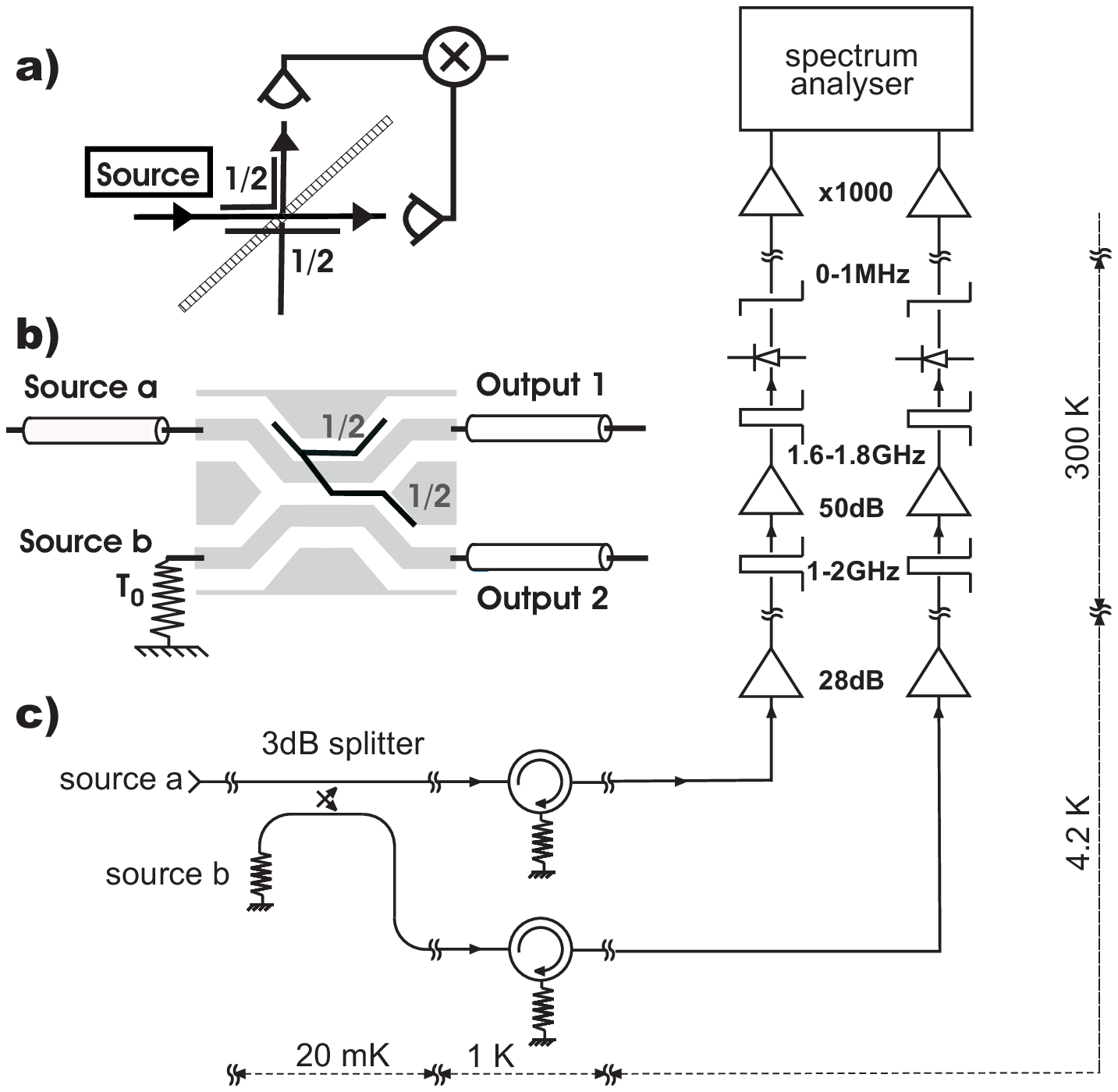}}\caption{{\small Schematics
of the HBT experiment (a), of the GHz beam splitter (b), of the
amplification and detection chains (c). A detailed description is
given in the text.}}\label{HBT}\end{figure}

The seminal experiment of Hanbury-Brown and Twiss (HBT)\cite{HBT},
consisted in two detectors correlating the power fluctuations of a
single electromagnetic source at the outputs of a beam splitter
(Fig.\ref{HBT}-a). Positive correlations observed for thermal
photons within the coherence time have been interpreted as
originating from photon bunching due to Bose-Einstein statistics
\cite{Purcell56}. Since correlations are very sensitive to
statistics, HBT experiments have been widely used to investigate
optical sources: laser light has poissonian fluctuations revealed
by vanishing HBT correlations; negative correlations have also
been reported for non-classical sources \cite{Grangier86}.

HBT correlations have also been observed with degenerate electrons
in quantum conductors, where the analogue of the photon flux is
the electrical current. Due to Fermi-statistics, electrons emitted
by contacts can be noiseless. As a result, sub-poissonian noise
and electron antibunching are observed
\cite{Reznikov95,Kumar96,Henny99,Oliver99}, in quantitative
agreement with predictions
\cite{Blanter00,Lesovik89,Buttiker90,Martin92}. Comparing photon
and electron statistics would remain a formal exercise unless one
notices the dual representation of the current as fermionic
excitations in a quantum \emph{conductor} and bosonic
electromagnetic modes in the external measuring \emph{circuit}.
This rises the intriguing question on how to relate the
sub-poissonian statistics of electrons fluctuations in the
conductor to the statistics of the emitted photons
\cite{Yurke84,Levitov93,Kinderman03}. A first theoretical answer
has been given in Ref.\cite{Beenakker01} where the photon
statistics is shown to remarkably deviate from the superpoissonian
black-body radiation. As shown below there is also a direct
relation between photon statistics and the fourth moment of the
electron Full Counting Statistics \cite{Kinderman03}.

But, how to measure reliably the photon statistics in quantum
circuits? Quantum effects in conductors, best displayed at
sub-Kelvin temperatures, require few photon number detection at
GHz frequencies and make experimental detection of TEM photon
fluctuations challenging. An elegant approach is provided by
Quantum Dots and Superconducting mesoscopic photon detectors
\cite{Aguado00,Deblock03}. However, a more versatile approach is
to use cryogenic low noise amplifiers (LNA) followed by detectors.

In this letter, we present the first HBT photon correlation
measurements in the GHz range and in the few photon population
number regime. The use of linear phase-insensitive amplification
prior to photon detection provides a new situation having no
optical counterpart. It is shown to preserve the nature of the HBT
correlations characterizing the photon source under test. The
sensitivity to various statistics is tested. First, an impedance
matched resistor is used which provides an ideal black-body
thermal source. Although expected, we show for the first time that
the Johnson-Nyquist noise is associated with superpoissonian noise
and positive HBT correlations characteristic of Bose-Einstein
thermal distribution. Second, a low noise radiofrequency generator
is used as a monochromatic coherent photon source. It is found to
generate coherent photons with poissonian statistics and vanishing
HBT correlations (like a Laser). Regarding auto-correlations,
which are sensitive to detection details, a quantum description of
the amplifier noise is made which quantitatively agrees with our
measurements.

Following Nyquist \cite{Nyquist28}, the current noise power of a conductor of conductance $G$
can be related to the power of TEM waves emitted in the external circuit. If $Z$ is the characteristic circuit
impedance, the average power in a small frequency range $\nu,\,\nu+d\nu$ is:
\begin{equation*}
\overline{P}=\frac{Z}{(1+GZ)^{2}}S_{I}(\nu)d\nu=\overline{N}h\nu d\nu
\end{equation*}
Here $N$ is the photon population, and
$S_{I}(\nu)=\overline{(\Delta I)^{2}}/d\nu$ the spectral density
of current fluctuation $\Delta I(t)$ filtered in the frequency
range $\nu,\,\nu+d\nu$ \cite{definition}. The fluctuations of the
current noise power entail \emph{fluctuations} in the photon power
at frequencies much lower than $\nu$. For classical currents, its
variance, proportional to the low frequency bandwidth $B$, is:
\begin{equation*}
\left\langle (\Delta P)^{2}\right\rangle
=\frac{Z^{2}}{(1+GZ)^{4}}\left\langle \left[(\Delta
I)^{2}(t)-\overline{(\Delta I)^{2}}\right]^{2}\right\rangle
\end{equation*}
and that of the photon population is directly related to the fourth moment of the current fluctuation:
\begin{equation*}
\overline{(\Delta N)^{2}}=\left( \frac{Ze^{2}}{h(1+GZ)^{2}}\right)
^{2}\times\frac{\overline{(\Delta I)^{4}}-\left[\overline{(\Delta
I)^{2}}\right]^{2}}{(e^{2}\nu d\nu)^{2}}
\end{equation*}
According to Glauber \cite{Glauber63}, thermal fluctuations of
a\textit{ classical }current lead to superpoissonian fluctuations
$\overline{(\Delta N)^{2}}=\overline{N}(1+\overline{N})$ or:
\begin{equation}\label{deltaP2}
\left\langle (\Delta P)^{2}\right\rangle
=2B\overline{N}(1+\overline{N})(h\nu)^{2}d\nu
\end{equation} with
$\overline{N}$ given by the Bose-Einstein distribution. In the
case of quantum conductors, the current is no longer a classical
observable and the unsymmetrized current noise operator
$\hat{S}_{I}(\nu
,t)\!=\!\int_{-\infty}^{+\infty}d\tau\,\hat{I}(t)\hat{I}(t+\tau)exp(i2\pi\nu
\tau)$ is needed \cite{Gavish00,Aguado00}. A full quantum
treatment can be found in Ref.\cite{Beenakker01} which shows that,
for conductors with large numbers of electronic modes, the
statistics of photons shows only small deviations from
Bose-Einstein distribution even when shot noise dominates over
thermal noise. However, a single electronic mode conductor with
transmission close to 1/2 is found to emit \emph{non-classical}
photons for frequency in the range $I/e$. Non-classical photons
are characterized by negative deviations $\overline{(\Delta
N)^{2}}-\overline{N}$ from the Poisson statistics, which is
precisely the quantity that HBT correlations measure. This
remarkable result and the connection with the fourth moment of the
full electron current statistics provide a strong motivation to
investigate HBT photon correlations in quantum circuits.

The radio-frequency equivalent of the optical HBT experiment is
shown in Fig.\ref{HBT}-b. Source (a) is the rf-photon source under
test. The emitted TEM photons propagate through a short 50 Ohm
characteristic impedance coaxial line and are fed to a cryogenic
3dB strip-line power splitter. The splitter scattering matrix,
measured with a network analyzer, is identical to that of an
optical separatrix (phases included). A built-in 50 Ohms resistor,
source (b), plays the role of the vacuum channel of the optical
case. Photon vacuum is achieved when its temperature $T_{0}$
satisfies $T_{0}\ll T_{Q}$ where $T_{Q}=h\nu/k_{B}$. Outputs (1)
and (2) of the power splitter are not immediately detected but a
1-2 GHz linear phase-insensitive amplification chain is inserted
before square-law detection (see Fig.\ref{HBT}-c). Each chain
consists in a microwave circulator followed by an ultra-low noise
cryogenic amplifier and room temperature amplifiers. The
circulators, at low temperature, ensure that amplifiers do not
send back photons towards (a) and (b). The detectors give an
output voltage proportional to the photon intensity \emph{after
amplification} $P_{1,2} ^{out}$. Their finite 1 $\mu \mathrm{s}$
integration time allows to monitor the low frequency photon
intensity fluctuations. A fast numerical spectrum analyser
calculates the autocorrelations $\left\langle (\Delta
P_{1,2}^{out} )^{2}\right\rangle $ and the HBT cross-correlations
$\left\langle \Delta P_{1}^{out}\Delta P_{2}^{out}\right\rangle $
in the band 40-200 kHz ($B$ =160 kHz).

In a first series of experiments, A and B, the sensitivity to
Bose-Einstein statistics is tested. In a third experiment C,
Poisson's statistics is tested using coherent photons. In A,
source (a) is a 50 conductor whose temperature $T$ is varied from
20 mK to several K, while source (b) realizes good photon vacuum
($T_{0}=17-20\,$mK$<T_{Q}$). A pair of 1.64-1.81 GHz filters
select a narrow band frequency around $\nu$ =1.72 GHz, with
$T_{Q}$ = 86 mK and $hd\nu\ll k_{B}T$. A quantum amplifier
description (see below) predicts the mean powers
$P_{i}=P_{i}^{out}/G_{i}$, referred to the input, and their
fluctuations:
\begin{eqnarray} \overline{P_{i}}=(\frac{\overline{N_{a}}}{2}+\frac{\overline{N_{b}}}{2}
+\frac{k_{B}T_{N,i}}{h\nu})h\nu d\nu\qquad i=1,&2&\label{Pmoy}\\
\left\langle (\Delta P_{i})^{2}\right\rangle
=2B(\frac{\overline{N_{a}}}{2}+\frac{\overline{N_{b}}}{2}+\frac{k_{B}T_{N,i}}{h\nu})\times\qquad\quad\quad&\quad&\nonumber\\
\times(\frac{1}{G_{i}}\frac{\overline{N_{a}}}{2}+\frac{\overline{N_{b}}}{2}
+\frac{k_{B}T_{N,i}}{h\nu})(h\nu)^{2}d\nu\approx\langle
P_{i}\rangle^{2}
\frac{2B}{d\nu}&\,&\label{auto}\\
\left\langle \Delta P_{1}\Delta P_{2}\right\rangle
=2B(\frac{\overline{N_{a}}}{2}-\frac{\overline{N_{b}}}{2})^{2}(h\nu)^{2}d\nu\quad&\quad&\label{corr}
\end{eqnarray}
Here $\overline{N_{a}}$ and $\overline{N_{b}}$ are the photon
populations of sources (a) and (b) given by Bose-Einstein
distribution, with $\overline {N_{b}}\approx0$ as $T_{0}<<T_{Q}$.
Autocorrelations differ from Eq.(\ref{deltaP2}) in two ways:
first, the factor 1 in the second parenthesis, revealing the
independent particle behavior of thermal photons is replaced by
$1/G_{i}$ ($G_{i}$ = 80 dB) when refered to the amplifier input;
secondly, an extra photon population is added due to amplifier
noise expressed in temperature units $T_{N,i}$ ($T_{N} \approx$
15K in the experiment A and $T_{N}\approx$ 6K in experiments B and
C). The HBT cross-correlation are \emph{unaffected} by
amplification.

\begin{figure} \centerline{\includegraphics{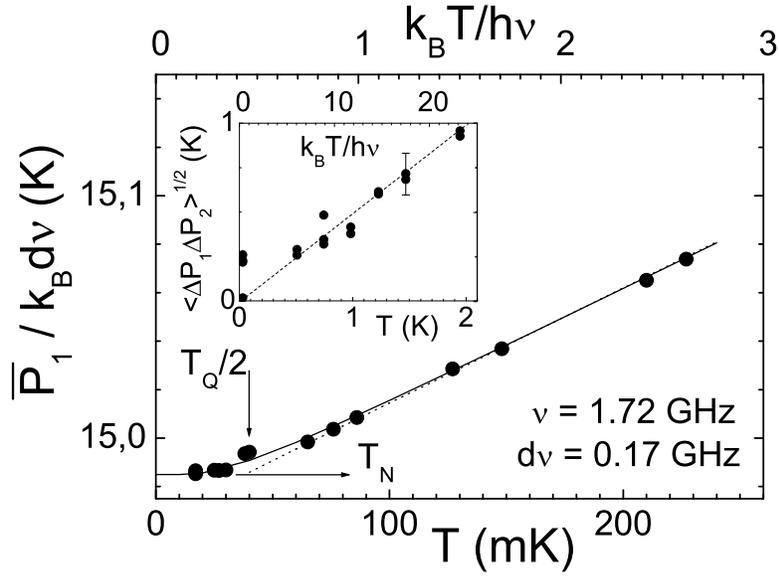}
}\caption{\small Thermal source at $\nu$ = 1.7 GHz. Main frame:
quantum crossover at $T_{Q}/2\approx$ 40 mK in the mean photon
power $\overline{P_{1}}$ as function of temperature $T$, or
effective occupation number $k_{B}T/h\nu$ for a beam splitter
temperature $T_{0}\approx$ 17 mK. The solid line is a
Bose-Einstein fit with an amplifier noise temperature
$T_{N}\approx$ 15 K. Inset: positive HBT correlations show the
bunching of thermal photons and the absence of residual
correlations at $T$ = 0 K. Experimental scatter lies within the
expected statistical uncertainty (error bar).}
\label{thermal20mK}\end{figure}

\begin{figure} \centerline{\includegraphics{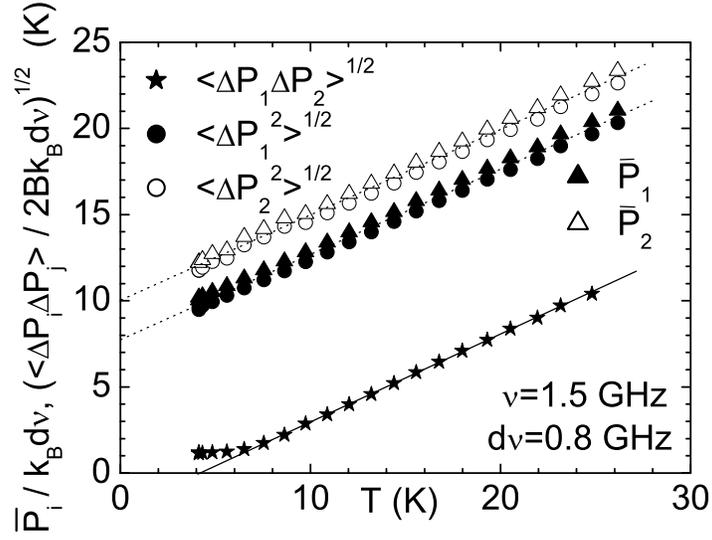}
}\caption{{\small : Superpoissonian correlations of a thermal
source in the classical regime: mean photon power
$\overline{P_{i}}$ and noise $\sqrt{\langle\Delta P_{i}\Delta
P_{j}\rangle}$ (bandwidth $B$ = 160 kHz) as function source
temperature $T$ with a beam splitter at $T_{0}\approx$ 4 K. The
temperature scale of $\sqrt{\langle(\Delta P_{i})^{2}\rangle}$ and
$\overline{P_{i}}$ is used as a calibration of the measuring line.
Cross-correlations $\sqrt {\langle\Delta P_{1}\Delta
P_{2}\rangle}$ are deduced without extra parameter and found to
vanish as $(T-T_{0})$ in agreement with Eq.(\ref{corr}). }}
\label{thermal4K}\end{figure}

Experimental results are shown in Fig.\ref{thermal20mK}. The solid
line is the Bose-Einstein theoretical fit of the mean photon power
with Eq.(\ref{Pmoy}) taking $T_{N}$ and $G$ as free parameters.
The experimental data reproduce well the quantum crossover at
$T_{Q}/2\approx$ 43 mK. This is the coldest quantum cross-over
ever reported for microwave photons. The parameter $G$ is
consistent with independent set-up calibration. The experimental
scatter, $\delta T\approx$ 1 mK in temperature units, corresponds
to the resolution $\delta T\approx2T_{N}/\sqrt{d\nu\Delta t}$
expected for few seconds acquisition time $\Delta t$. This
unprecedented sensitivity ammounts to flux rate variations of 1
photon per $\mu \mathrm{s}$. The HBT cross-correlations
$\langle\Delta P_{1}\Delta P_{2}\rangle$ are expected to vary like
$\left( \overline{N_{a}}\right) ^{2}\approx\left(  T/T_{Q}\right)
^{2}$ for $T>T_{Q}$ as $\overline{N_{B}}=0$. This is exactly what
we observed in Fig.\ref{thermal4K}. This provides the first
evidence for Bose-Einstein correlations of photon emitted by a
resistor in the few photon number limit at sub-Kelvin temperature.
The correlation resolution is $\delta\left\langle (\Delta
P)^{2}\right\rangle \approx\left\langle (\Delta
P_{1,2})^{2}\right\rangle /\sqrt{B\Delta t}$. This corresponds to
detect population fluctuations of $\sqrt{\overline{\Delta
N_{1}\Delta N_{2}}}\approx k_{B}T_{N}/\left(h\nu\sqrt[4]{B\Delta
t}\right)\approx$ 1.5 for our experimental $\Delta t$ =1000 s.

Experiment B provides evidence for vanishing HBT correlations
$\langle\Delta P_{1}\Delta P_{2}\rangle\approx\left(
\overline{N_{a}}-\overline{N_{b}}\right)^{2}$ when
$\overline{N_{a}}=\overline{N_{b}}$. The experiment is performed
at higher temperature ($T$ = 4-24 K, $T_{0}$ = 4 K). Source (b) is
no longer in the ground state: $\overline{N_{b}}(\nu)\approx
k_{B}T_{0}/h\nu\approx40$. Here, the central frequency is $\nu=
1.5$ GHz, the bandwidth $d\nu\approx$ 0.8 GHz and the amplifier
noise temperatures $T_{N,i}=6\,\mathrm{K}$ and 8 K. As shown in
Fig.\ref{thermal4K}, $\overline{P_{1,2}}$ linearly depend on
temperature in agreement with Eq.(\ref{Pmoy}). The $T$ =
0-extrapolates define the LNA noise temperatures. In our analysis
the linear slopes give $G_{1}$ =1.53 $10^{8}$, ($\Delta\nu_{1}$ =
0.75 GHz), $G_{2}$ = 1.17 $10^{8}$ ($\Delta\nu_{2}$ = 0.88 GHz) in
fair agreement (within 20 \%) with independent calibration. The
square root of the auto-correlation shows linear variation with
temperature exemplifying the super-poissonian noise of thermal
photons given by Eq.\ref{auto}. There are no free parameter left
for the HBT correlations as gains are known. They are found
positive and quantitatively agree with Eq.(\ref{corr}) (solid line
in Fig.\ref{thermal4K}). They \emph{vanish} for equal temperature
sources.

\begin{figure}
\centerline{\includegraphics{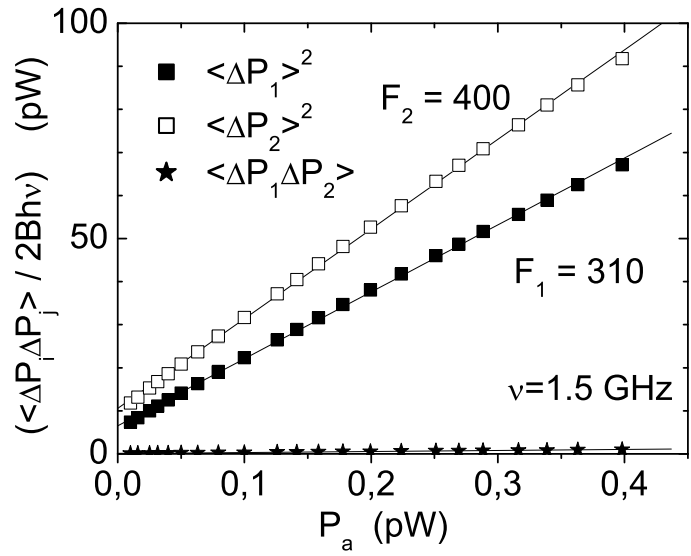}} \caption{{\small
Poissonian correlations of a coherent monochromatic source.
Autocorrelations are linear in the input RF power with a large
Fano factor $F$ due to amplification and attenuation. Cross
correlations vanish in accordance with theoretical predictions.}}
\label{coherent}
\end{figure}

Experiment C tests the sensitivity to a different statistics. A
microwave source of frequency $\nu_{0}$ = 1.5 GHz and bandwidth
$\approx$ 100 Hz generates monochromatic photons. Its output is
attenuated at cryogenic temperature $T_{att}$. The average power
$\overline{P_{a}}$ in (a), is chosen comparable to that delivered
by the thermal source in experiment B. As for a Laser, the source
is expected to generate coherent photon states \cite{Glauber63}.
The photon statistics being Poissonian, the low frequency power
fluctuations of source (a) in bandwidth $B$ are $\left\langle
(\Delta P_{a})^{2}\right\rangle =2Bh\nu_{0}\overline{P_{a}}$. An
important question is whether attenuation and amplification before
detection change the statistics and if yes how? According to
references \cite{Caves82,Yamamoto86}, the commutation rules for
bosonic input and output operators imply the addition of an extra
bosonic operator describing the amplifier/attenuator noise.
Applying this quantum constraint, we find (in units referred to
the amplifier chain input):
\begin{eqnarray*}
\left\langle (\Delta P_{1,2})^{2}\right\rangle
&=&2Bh\nu_{0}F_{1,2} \overline{P_{1,2}}    \qquad\left\langle
\Delta P_{1}\Delta P_{2} \right\rangle =0\\ F_{1,2}
&=&1+2k_{B}\frac{T_{att}+T_{N1,2}}{h\nu_{0}}
\end{eqnarray*}
where we have used $T_{att}=T_{0}$. Again, amplification has
\emph{no effect} on HBT cross-correlations and the absence of
cross correlation characterizing Poisson's statistics remains. In
the autocorrelations, a Fano factor $F$ appears due to
amplification noise. The results of experiment C are shown on
Fig.\ref{coherent}. Cross- and auto-correlations are plotted
versus detected power referred to the input. Residual cross
correlations, due to to small set-up imperfections, remain
negligible at the scale of the autocorrelations. Auto-correlations
are perfectly linear with power as expected. We measure a large
Fano factor $F_{1,2}$ = 310, 400 in accordance with the order of
magnitude of $(T_{att}+T_{N1,2})/h\nu_{0}$.

In conclusion, the highly sensitive HBT photon correlations
performed in the GHz range at subKelvin temperature using phase
insensitive LNA and square-law detection easily discriminate
between different statistics. In particular the cross-correlations
are unaffected by amplification details. The method, simple,
versatile and based on currently available electronics, can be
easily reproduced in other laboratories. It appears to be very
suitable to study the photon population statistics of TEM modes
emitted by quantum conductors or equivalently the fourth moment of
current fluctuations.

\begin{acknowledgments}
We thank B. Lazareff and J.Y. Chenu, from IRAM, for providing us
with cryogenic amplifiers. The laboratoire Pierre Aigrain is
"unit{\'e} mixte de recherche" (UMR 8551) of the Ecole Normale
Sup{\'e}rieure, the CNRS and the Universities Paris 6 and Paris 7.
\end{acknowledgments}

\newpage

\end{document}